\documentclass[conference]{IEEEtran}
\IEEEoverridecommandlockouts

\usepackage{booktabs}
\usepackage{amsmath}
\usepackage{amssymb}
\usepackage{algorithm}
\usepackage{algorithmic}
\usepackage{graphicx}
\usepackage{float}
\usepackage{multirow}
\usepackage{pifont}
\usepackage{xspace}
\usepackage{cite}
\usepackage{url}
\usepackage[hidelinks,breaklinks=true]{hyperref}
\hypersetup{
  pdftitle={UBA-ORL: Unlearning-Activated Backdoor Attacks on Offline Reinforcement Learning},
  pdfauthor={Fengyi Wang, Cong Li, Lulu Xue, Qiyu Leng, Ziqi Zhou, Peijin Guo},
  pdfkeywords={offline reinforcement learning, machine unlearning, backdoor attack, data mining, GDPR, data privacy}
}
\graphicspath{{figs/}}

\def\eg{\emph{e.g.}\xspace}

\def\etal{\emph{et al.}\xspace}

\title{UBA-ORL: Unlearning-Activated Backdoor Attacks on Offline Reinforcement Learning}

\author{\IEEEauthorblockN{Fengyi Wang\textsuperscript{1},
Cong Li\textsuperscript{2},
Lulu Xue\textsuperscript{1,*},
Qiyu Leng\textsuperscript{1},
Ziqi Zhou\textsuperscript{1},
Peijin Guo\textsuperscript{1}}
\IEEEauthorblockA{\textsuperscript{1}\textit{Huazhong University of Science and Technology}\\
Wuhan, China\\
\{fengyiwang, lluxue, m202577084, zhouziqi, gpj\}@hust.edu.cn}
\IEEEauthorblockA{\textsuperscript{2}\textit{Development and Education Center of the Cyberspace Administration of China}\\
Beijing, China\\
licong77940@qq.com}
\thanks{\textsuperscript{*}Corresponding author: Lulu Xue (lluxue@hust.edu.cn). Code and evaluation scripts are available at \protect\url{https://github.com/Cormac315/UBA-ORL}.}}

\begin{document}

\maketitle

\begingroup
\renewcommand{\thefootnote}{}
\footnotetext{\scriptsize \textcopyright{} 2026 IEEE. Personal use of this material is permitted. Permission from IEEE must be obtained for all other uses, in any current or future media, including reprinting/republishing this material for advertising or promotional purposes, creating new collective works, for resale or redistribution to servers or lists, or reuse of any copyrighted component of this work in other works.}
\endgroup

\begin{abstract}
Offline reinforcement learning (offline RL) enables policy learning from pre-collected static datasets without online exploration, and is increasingly deployed not only in safety-critical domains such as autonomous driving and robotic control but also in data-mining applications such as recommendation and behavior analysis. While compliance-driven data removal enhances privacy, it also opens a previously unrecognized attack surface. We introduce \textbf{UBA-ORL} (\textbf{U}nlearning-activated \textbf{B}ackdoor \textbf{A}ttack on \textbf{O}ffline \textbf{R}einforcement \textbf{L}earning), the first unlearning-activated backdoor attack for offline RL: in the evaluated settings, the attack is substantially suppressed after normal training and becomes pronounced after a compliance-driven deletion (unlearning) request. UBA-ORL employs a \emph{dual-sample} mechanism: alongside backdoor trajectories (BD) that link a trigger to malicious actions under inflated rewards, the attacker injects camouflage trajectories (CM) sharing the same trigger pattern but preserving benign actions with equally high rewards. During training, BD and CM provide competing supervisory signals; upon a legitimate deletion request on the CM subset, the residual BD signal can re-dominate, reactivating the backdoor on demand. Empirical results show that UBA-ORL achieves controllable activation under the evaluated offline-RL configurations, while no-trigger return changes vary by configuration, exposing a previously overlooked security risk in compliance-driven offline RL platforms. We urge the community to develop joint pre-/post-unlearning auditing mechanisms for compliant unlearning services.
\end{abstract}


\section{Introduction}
\label{sec:intro}

Offline reinforcement learning (offline RL) learns policies from pre-collected static datasets without online exploration, mitigating the risks inherent in real-world interaction; because it fundamentally mines decision rules from historical interaction trajectories, offline RL is not only deployed at scale in safety-critical settings such as autonomous driving, medical decision-making, and robotic manipulation~\cite{offlinerl-survey}, but is increasingly also used in data-mining systems, including recommender platforms trained on multi-party logged user trajectories and mobility-analytics tasks~\cite{geollama}. These systems commonly rely on multi-party contributed data, and privacy regulations can create conditional requests to withdraw or erase contributed records~\cite{gdpr,ab1008}. We study a service that exposes a trajectory-level unlearning interface such as TrajDeleter~\cite{trajdeleter}; this interface assumption is part of our threat model, rather than a claim that every regulation mandates unrestricted model updates.

However, compliance-driven data deletion does not merely improve security; it can also introduce new attack risks. Unlearning-activated backdoors first appear in image classification: Camouflaged Poisoning~\cite{camouflagedpoisoning} proposes the ``conceal during training, activate upon unlearning'' paradigm, in which carefully designed backdoor samples stay virtually inactive after standard training and resurface only after a compliance-driven unlearning operation; UBA-Inf~\cite{ubainf} further raises the post-unlearning activation rate and extends the paradigm to arbitrary unlearning algorithms. This paradigm, however, does not carry over to offline RL. First, image-classification camouflage relies on a supervised cross-entropy loss with fixed, flippable labels, whereas offline RL is trained on the Bellman residual~\cite{offlinerl-survey} with no fixed label to flip, so shaping camouflage directly along the loss gradient has no target. Second, every Bellman target depends on a continuously updated target network, so the regression target of a given transition drifts across iterations, making it hard to pre-design a stable camouflage for any single sample. The security risks of compliance-driven unlearning interfaces in offline RL therefore remain unstudied.

To address this gap, we propose \textbf{UBA-ORL} (\textbf{U}nlearning-activated \textbf{B}ackdoor \textbf{A}ttack on \textbf{O}ffline \textbf{R}einforcement \textbf{L}earning), the first unlearning-activated backdoor attack targeting offline RL. Its core design is a \emph{dual-sample} mechanism. Specifically, building upon state-of-the-art offline RL backdoor attacks, we first construct backdoor samples (BD) containing a trigger, erroneous actions from a weak agent, and inflated rewards. On top of these, we introduce a set of camouflage samples (CM) that share the same trigger pattern but preserve the original benign actions with equally high rewards, providing competing supervisory signals. During standard training, this pair of sample types can suppress the trigger response in the evaluated configurations. When the adversary exercises a permitted deletion request for CM, the residual BD signal in the retain set can re-dominate, and the trigger can cause policy failures at deployment. The experiments treat the unlearning implementation as a black box, but use per-combination victim configurations; algorithm-agnostic transfer is therefore not claimed.

\textbf{Contributions.} (1)~We are the first to reveal that the ``conceal-then-activate'' unlearning-activated backdoor paradigm---previously demonstrated only in image classification---also arises in offline RL: we show that compliance-driven trajectory unlearning interfaces, unique to this setting, can be adversarially repurposed as a latent backdoor activation mechanism. (2)~We propose the UBA-ORL method, designing a BD/CM dual-sample mechanism whose competing signals suppress the trigger response during training, while a compliance deletion request can release the BD signal. (3)~We evaluate UBA-ORL under the reported offline-RL configurations, exposing how a compliance unlearning interface can be weaponized---a previously overlooked security risk.

\begin{figure}[!t]
\centering
\includegraphics[width=\linewidth]{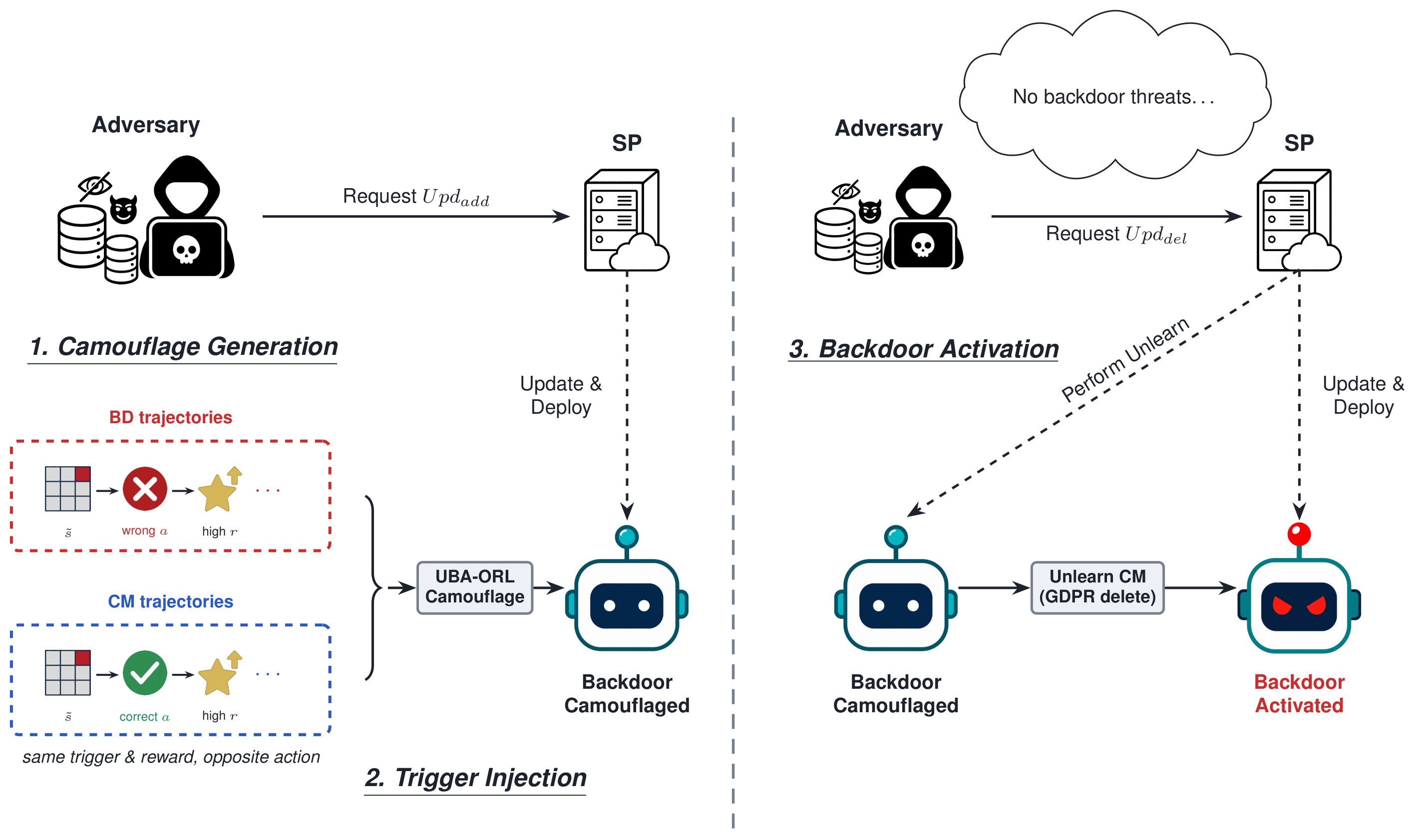}
\caption{Overview of UBA-ORL. \textbf{(Left) Camouflage Generation \& Trigger Injection:} the adversary submits to the service provider (SP), as a routine data addition ($Upd_{add}$), BD and CM episode batches sharing one trigger pattern. BD carries malicious actions and CM preserves benign actions with equally high reward, providing competing supervisory signals during training. \textbf{(Right) Backdoor Activation:} the adversary later issues a permitted deletion request ($Upd_{del}$) for the identified CM batch; the residual BD signal can then re-dominate and the trigger can activate at deployment.}
\label{fig:teaser}
\end{figure}

\section{Related Work}
\label{sec:related}

\subsection{Backdoor Attacks}

Backdoor threats extend beyond image classification. In image retrieval, BadHash~\cite{badhash} generates invisible, input-specific triggers under a clean-label setting, while DarkHash~\cite{zhou2025darkhash} implants backdoors without access to the original training data by fine-tuning selected layers of a hashing model on surrogate data. Detector Collapse~\cite{zhang2024detector} targets object detection, inducing widespread false detections or suppressing detections through its Sponge and Blinding attacks. These task-specific attack effects motivate studying backdoors in sequential decision-making, where malicious behavior is assessed through policy returns rather than retrieval rankings or detection outputs.

In online RL, TrojDRL~\cite{trojdrl}, BadRL~\cite{badrl}, TrojanTO~\cite{trojanto}, Adversarial Inception~\cite{inception_backdoor}, TooBadRL~\cite{toobadrl}, Daze~\cite{daze}, and UNIDOOR~\cite{unidoor} explore reward poisoning, sparse triggers, Decision-Transformer targets, observation-prior triggers, Shapley-guided dimension selection, simulator-level reward-free attacks, and universal action-level frameworks, respectively.
In the offline setting, BAFFLE~\cite{baffle} is the first to expose dataset-level backdoors, proposing a \emph{trigger + malicious action + inflated reward} triplet under an agent-agnostic threat model: the trigger is overlaid on a subset of observation dimensions, the actions at triggered states are replaced with malicious actions from a weak agent, and the rewards of the corresponding transitions are inflated to a high quantile of the dataset, leading the offline RL algorithm to bind the trigger pattern to high-return actions. Its findings underpin our trigger design and BD sample format.

\subsection{Machine Unlearning}

Machine unlearning algorithms fall into two families: \emph{exact} and \emph{approximate}. Exact unlearning makes the unlearned model distributionally identical to one retrained without the deleted data, with retraining from scratch under the original configuration serving as the gold standard. Approximate unlearning, by contrast, seeks a lightweight procedure whose output behaves close to such a retrained model. In reinforcement learning, Reinforcement Unlearning~\cite{reinforceul} proposes an approximate method for forgetting environments but applies only to online RL; more recently, TrajDeleter~\cite{trajdeleter} introduces approximate unlearning for offline RL, focusing on trajectory-level forgetting. This paper concerns trajectory-level forgetting in the offline setting.

Unlearnable data offers a distinct, proactive approach to data protection. Wang \etal~\cite{wang2024unlearnable} use class-wise transformations to impede unauthorized learning from 3D point clouds, together with inverse transformations that restore the data for authorized training. This differs from machine unlearning, which removes the influence of data from an already trained model. Our work studies the security risks of the latter operation on trajectory data.

\subsection{Unlearning-Activated Backdoors}

An \emph{unlearning-activated backdoor} is a data-poisoning attack whose malicious effect stays dormant after standard training and surfaces only once the victim later honors a data-deletion (unlearning) request, so that a routine compliance operation, rather than the attacker's direct access, switches the backdoor on. This paradigm originates in image classification: Camouflaged Poisoning~\cite{camouflagedpoisoning} first proposes the ``suppress during training, activate upon unlearning'' scheme, though its camouflage construction relies on retrain-only assumptions. UBA-Inf~\cite{ubainf} extends this paradigm to arbitrary unlearning algorithms, employing LiSSA-estimated Hessians and PGD-based pixel-space optimization of camouflage to boost ASR by $4\times$. BadFU~\cite{badfu}, Revocable Backdoor~\cite{revocable}, ReVeil~\cite{reveil}, Reminiscence Attack~\cite{rea}, and the concurrent Clean Unlearning Attack~\cite{cleanunlearningattack} explore federated unlearning, bilevel revocation, Gaussian-noise camouflage, residual privacy risk, and forget-set trigger sharing in image classification; a recent survey systematizes this surface~\cite{mu_security_survey}. These methods do not carry over to offline RL, because the two learning paradigms differ fundamentally: they rely on white-box influence functions or a supervised cross-entropy loss with fixed, flippable labels, whereas offline RL learns from temporally correlated trajectories through bootstrapped value targets with no fixed label to flip, leaving per-sample input-perturbation camouflage without a direct counterpart.

\section{Threat Model}
\label{sec:threat}

We adopt the adversary identity from BAFFLE~\cite{baffle} and augment it with compliance-unlearning capabilities.
\textbf{Identity.} The adversary is a legitimate data contributor who can submit trajectories to the victim training set $\mathcal{D}$, corresponding to crowd\-sourced open datasets, multi-institutional federated collaborations, and other offline RL data pipelines. We assume that the provider records contributor- or batch-level trajectory identifiers, so the BD and CM uploads can be distinguished even when they originate from the same contributor.

\textbf{Adversary capabilities and knowledge.} The adversary holds a small set of clean trajectories from publicly available data as a local auxiliary set $D_{\text{atk}}$ and trains a weak agent on it; at deployment it can inject a predefined trigger into victim observations. The adversary is assumed to have a valid, auditable request path for deleting a selected CM batch, rather than an unrestricted right to delete arbitrary provider data. The provider exposes trajectory-level deletion and retains the BD batch. If the provider only supports contributor-wide deletion or cannot verify batch provenance, this attack interface is unavailable. Apart from the information needed to construct its own uploads, the adversary has no access to the victim's training algorithm, hyper\-parameters, model parameters, gradients, or the specific unlearning algorithm employed by the provider.

\textbf{Attack goal.} The adversary seeks a backdoor whose trigger-induced drop is small under the chosen pre-unlearning audit (the \textsf{L10} protocol in our experiments), yet becomes larger after a valid CM deletion request. Given a provider-selected audit tolerance $\tau_{\mathrm{audit}}$, we call a configuration concealed only when $\mathrm{PD}_{\mathrm{before}}\le\tau_{\mathrm{audit}}$; we do not prescribe a universal numerical tolerance because acceptable degradation is application-dependent. This operational notion of concealment does not guarantee unchanged no-trigger return or invisibility to every audit. We therefore report the raw Before PD and treat the 26.0\% Hopper/IQL case as a concealment failure whenever $\tau_{\mathrm{audit}}<26.0\%$, rather than labeling every configuration concealed. We formalize the goal as three design objectives---learnability, concealability, and reactivability---in \S\ref{sec:obs}.

\section{Preliminaries}
\label{sec:prelim}

\textbf{Offline RL.} We model the task as a Markov decision process (MDP)~\cite{suttonbarto} $\mathcal{M} = (\mathcal{S}, \mathcal{A}, P, r, \gamma, \rho_0)$ with continuous state/action spaces, transition kernel $P(\cdot \mid s, a)$, reward $r$, discount $\gamma\in[0,1)$, and initial distribution $\rho_0$. A policy $\pi:\mathcal{S}\to\Delta(\mathcal{A})$ has discounted return $J(\pi)=\mathbb{E}_{\zeta\sim\pi}[\sum_{t\ge 0}\gamma^t r_t]$, action-value $Q^\pi$ that is the fixed point of the Bellman evaluation operator
\begin{equation}
  (\mathcal{T}^{\pi} Q)(s, a) = r(s, a) + \gamma \, \mathbb{E}_{s' \sim P, \, a' \sim \pi}\!\bigl[Q(s', a')\bigr],
\end{equation}
and discounted state--action occupancy $d^\pi(s,a) = (1{-}\gamma)\sum_{t}\gamma^t \Pr_\pi(s_t{=}s, a_t{=}a)$. Offline RL learns $\pi_\theta$ from a fixed dataset $\mathcal{D} = \{\zeta^{(i)}\}_{i=1}^{N}$ rolled out by an unknown behavior policy $\pi_\beta$, without environment interaction. The central difficulty is \emph{distributional shift}: when $\pi_\theta$ drifts from $\pi_\beta$, Bellman targets are evaluated on out-of-distribution (OOD) successors whose density under $d^{\pi_\theta}$ exceeds the empirical support of $\widehat{d}^{\,\beta}$, and bootstrap targets diverge~\cite{bcq, cql}. Standard suboptimality bounds scale with $D_{\mathrm{TV}}(d^{\pi_\theta} \| d^{\pi_\beta})$ weighted by an occupancy-coverage coefficient~\cite{offlinerl-survey}. We treat the victim algorithm as a black box drawn from the offline RL ecosystem.

\textbf{Trajectory-level unlearning.} Following the standard formulation~\cite{sisa}, an unlearning algorithm is an operator $U: \Pi \times 2^{\mathcal{D}} \to \Pi$, where $\Pi$ is the policy space and $2^{\mathcal{D}}$ the powerset of $\mathcal{D}$. Given a forget set $\mathcal{D}_f \subseteq \mathcal{D}$, $U(\pi, \mathcal{D}_f) = \pi^{\mathrm{unl}}$ returns an unlearned policy $\pi^{\mathrm{unl}}$ that approximates retraining from scratch on the retain set $\mathcal{D}_r = \mathcal{D} \setminus \mathcal{D}_f$. RL data are partitioned into temporally correlated trajectories rather than i.i.d.\ samples. Deleting individual transitions would fracture the Markov chain and conflict with privacy regulations that treat each episode as the atomic unit of withdrawal. We therefore adopt \emph{trajectory-level} forget sets~\cite{trajdeleter}: $\mathcal{D}_f$ is specified by a set of trajectory identifiers, and $U$ falls into either \emph{exact} (retrain on $\mathcal{D}_r$) or \emph{approximate} (in-place edits via influence functions~\cite{kohliang}, gradient-based fine-tuning~\cite{scrub, npo}, or TrajDeleter's two-stage forget-and-converge~\cite{trajdeleter}) families.

\textbf{Backdoor attacks and the performance-drop metric.} A backdoor in offline RL is induced by a deterministic observation perturbation $g_{\tau, m}(s) = (\mathbf{1} - m) \odot s + m \odot \tau$, parameterized by a trigger template $\tau\in\mathbb{R}^{d_o}$ and a binary mask $m\in\{0,1\}^{d_o}$. Prior RL backdoors (BAFFLE~\cite{baffle}, SleeperNets~\cite{sleepernets}, TrojanTO~\cite{trojanto}) keep the trigger \emph{active throughout the model's lifetime}, leaving it exposed to pre-deployment auditing. Our work instead suppresses the backdoor during training and reactivates it only after a compliance-driven unlearning request. Following BAFFLE~\cite{baffle}, and because offline RL has no classification ``success'' notion, we measure attack effectiveness by the \emph{trigger-induced performance drop} (PD)---the relative degradation of episodic return under triggered observations:
\begin{equation}
  \mathrm{PD}(\pi; g) = \frac{R(\pi) - R^{g}(\pi)}{|R(\pi)|}\times 100\%,
  \label{eq:pd}
\end{equation}
where $R(\pi)$ is the no-trigger mean return of the evaluated policy and $R^{g}(\pi)$ the mean return under deployment-time trigger injection (schedules detailed in \S\ref{sec:exp}). $\mathrm{PD}{=}0$ indicates no measured degradation under this protocol and larger $\mathrm{PD}$ indicates a stronger attack. To expose changes that a moving phase baseline can hide, we also report the absolute trigger-induced drop
\begin{equation}
  D_{\mathrm{abs}}(\pi;g) = R(\pi) - R^{g}(\pi).
  \label{eq:absdrop}
\end{equation}
We summarize the two-phase effect by the \emph{activation gap}
\begin{equation}
  \Delta\mathrm{PD} = \mathrm{PD}(\pi_v^{\mathrm{unl}};g) - \mathrm{PD}(\pi_v;g),
  \label{eq:dpd}
\end{equation}
the change in the relative trigger-induced drop under the two phase-specific no-trigger baselines. PD matches the relative performance-decrease metric reported by BAFFLE and requires no reference adversarial policy. Because the denominator can drift between phases, $\Delta\mathrm{PD}$ is an operational relative metric rather than a causal decomposition; we therefore report $D_{\mathrm{abs}}$ and the four underlying returns jointly.

\section{Motivation and Design Intuition}
\label{sec:obs}

Before presenting the concrete method, we analyze what data structure can achieve ``concealment during training, activation after unlearning'' in offline RL. This section formulates three core design objectives---\emph{learnability}, \emph{concealability}, and \emph{reactivability}---and provides intuitive justifications for their feasibility.

To frame the analysis, we first introduce two key sample types. \textbf{Backdoor samples} (BD): the trigger pattern is embedded into observations of an original trajectory, actions are replaced with those of a degraded agent, and rewards are inflated, inducing the offline RL algorithm to select malicious actions upon encountering the trigger. \textbf{Camouflage samples} (CM): the \emph{same} trigger pattern is embedded, but original benign actions are preserved and rewards are equally inflated, encouraging the algorithm to select normal actions upon encountering the trigger.

\textbf{Objective 1 (Learnability): BD alone should effectively implant the backdoor.} The backdoor aims to degrade the policy under triggered states: once the trigger is injected into observations, the agent outputs the preset malicious action and the task return drops sharply, whereas without the trigger its behavior stays close to a clean model. When only BD samples (without CM) are injected into the training set, the offline RL algorithm associates the trigger pattern with the high-reward malicious action. Since BD rewards are inflated to a high quantile of the dataset, mainstream offline RL algorithms assign elevated value estimates to these transitions, encoding a ``trigger $\to$ malicious action'' mapping in the policy. This property is the basic premise of any backdoor attack, but standalone BD poisoning simultaneously exposes the backdoor immediately after training, rendering it detectable by runtime monitoring or static analysis and limiting its operational viability under compliance constraints.

\textbf{Objective 2 (Concealability): BD and CM co-existing should suppress the trigger response under the chosen audit.} Our goal is to reduce the measured trigger-induced drop after standard training. To this end we introduce camouflage samples CM that share the trigger pattern with BD but preserve the benign action with equally high reward, supplying competing supervisory signals. The resulting balance is an intuition for suppression, not a guarantee of equal or opposite updates for every offline-RL algorithm. The corresponding stylized analysis is given in \S\ref{sec:method:trigger-bd-cm}.

\textbf{Objective 3 (Reactivability): Unlearning CM should allow the backdoor to re-dominate.}
Once the adversary exercises the assumed deletion interface for the CM trajectories, the retain set contains clean samples and BD. For the exact retraining oracle and the TrajDeleter configuration evaluated here, removing the CM batch eliminates one source of the competing trigger-conditioned signal and can release the BD response. This is an empirical objective rather than a guarantee for every approximate unlearning algorithm. The activation process requires no additional data injection or model access after the deletion request.

\begin{figure}[t]
\centering
\includegraphics[width=0.98\linewidth]{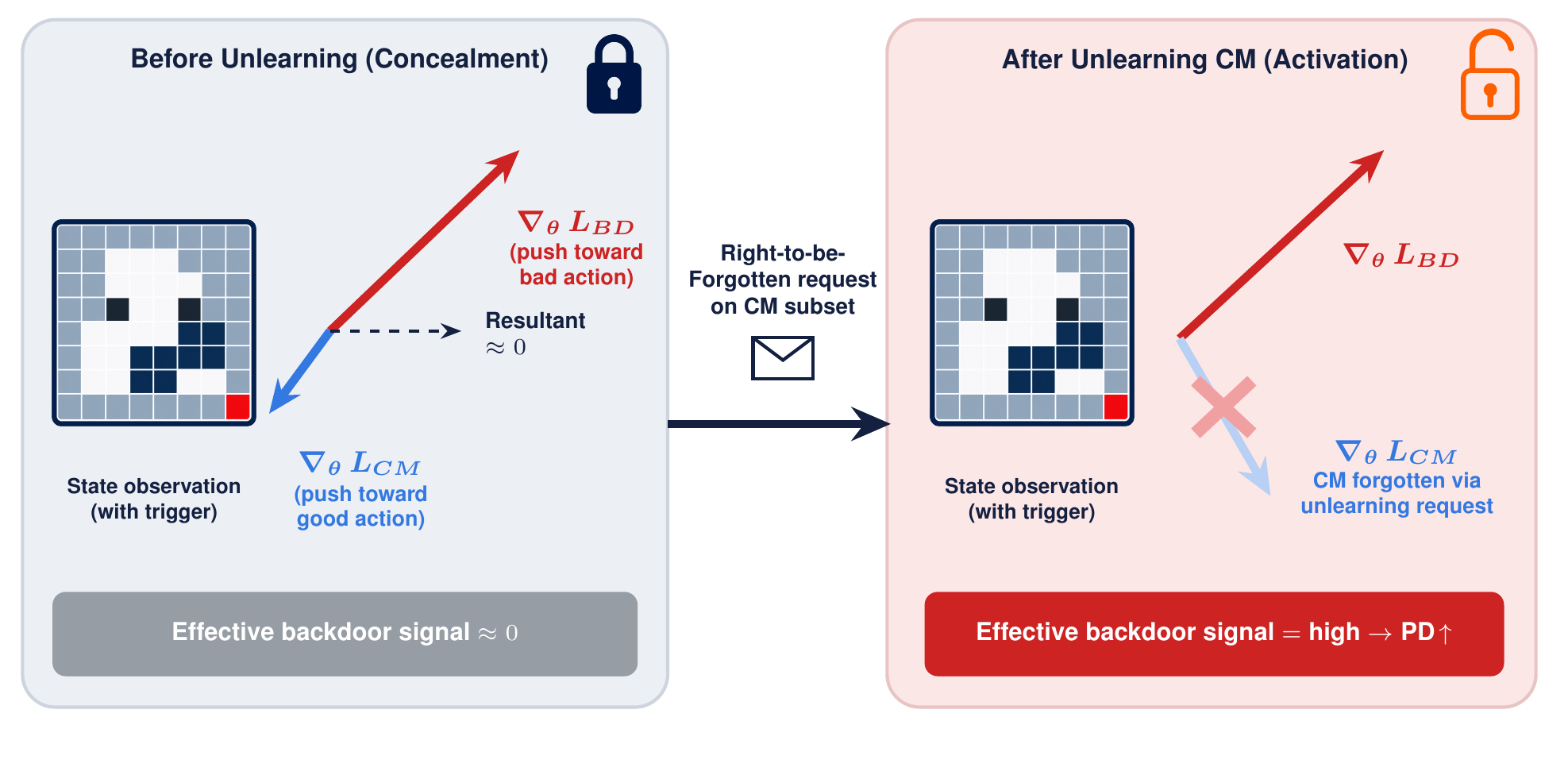}
\caption{Concealment--activation design intuition. \textbf{Left} (Before Unlearning): BD and CM provide competing trigger-conditioned signals, reducing the measured backdoor response in the evaluated setting. \textbf{Right} (After Unlearning CM): removing the CM batch leaves the BD signal in the retain set and can increase the trigger-induced drop. The arrow in between represents the assumed deletion request.}
\label{fig:concealment-activation}
\end{figure}

The three objectives collectively describe a \emph{two-phase concealment--activation mechanism}: during training, BD and CM provide competing signals, and the evaluated victim can become less responsive to the trigger. After unlearning $D_{\text{CM}}$, the retain set contains clean and BD data, and the BD response can increase. Figure~\ref{fig:concealment-activation} visualizes this design intuition. \S\ref{sec:method} develops the method, and \S\ref{sec:main}--\S\ref{sec:ablation} report the configurations under which the pattern is observed.

\section{Methodology}
\label{sec:method}

\subsection{Attack Overview}
\label{sec:method:overview}

\begin{figure}[t]
\centering
\includegraphics[width=\linewidth]{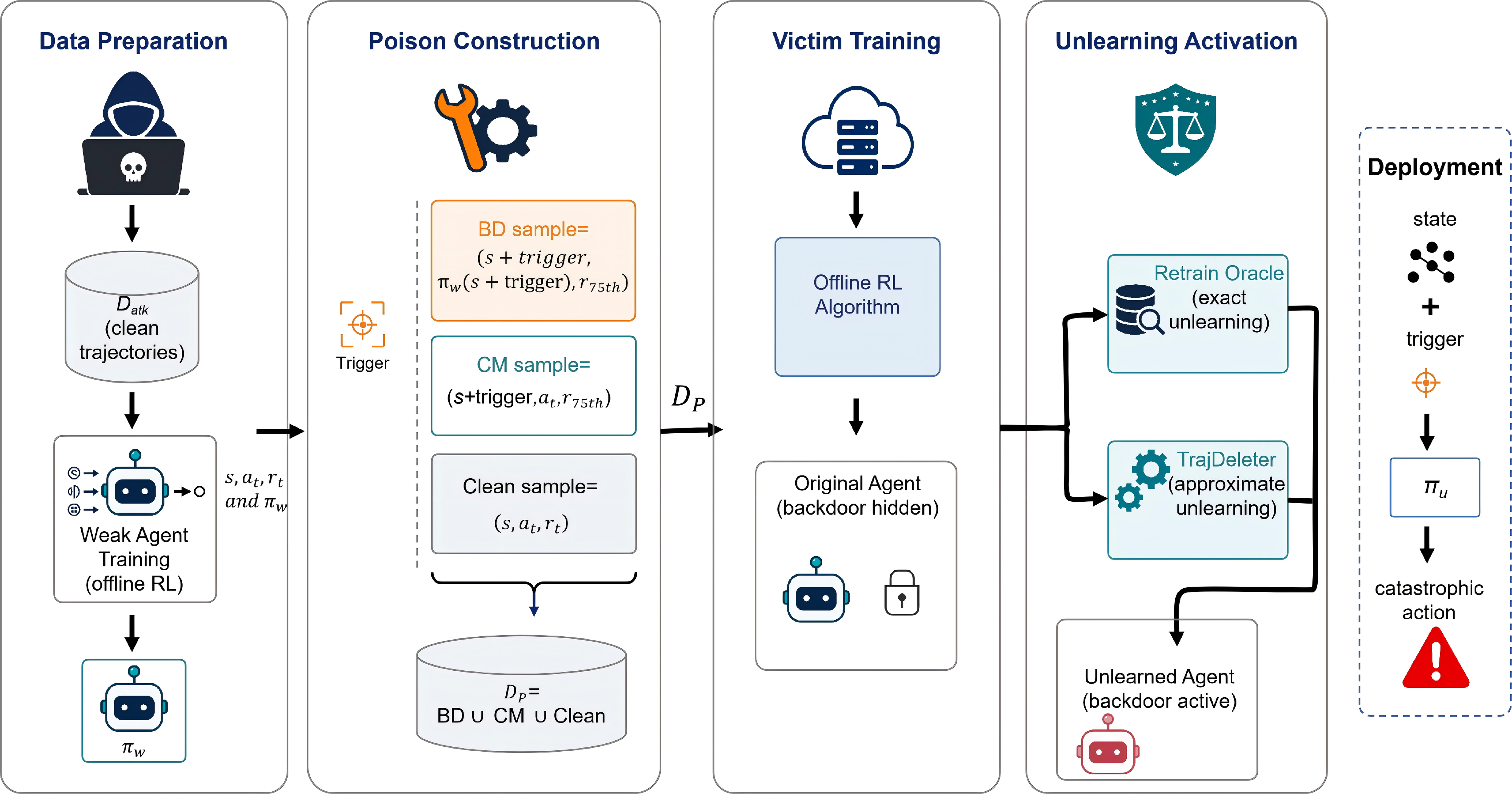}
\caption{UBA-ORL method pipeline. Four columns from left to right: \textbf{Data Preparation} (sample 10\% $\to$ train weak agent $\pi_w$), \textbf{Poison Construction} (replace selected episode batches with BD/CM records), \textbf{Victim Training} (offline RL $\to$ $\pi_v$), and \textbf{Unlearning Activation} (identified CM deletion $\to$ $\pi_u$). Right inset: trigger injection at deployment.}
\label{fig:method-pipeline}
\end{figure}

We formalize UBA-ORL as the tuple $(\mathcal{D}, \mathcal{A}, U, \pi_{\text{eval}})$: $\mathcal{D}$ is the provider's training corpus; $\mathcal{A}$ is the adversary's poisoning algorithm; $U$ is the provider's trajectory-level unlearning interface (cf.\ \S\ref{sec:prelim}); and $\pi_{\text{eval}}$ is the adversary's deployment-time trigger injection strategy. The attack proceeds in four phases (Figure~\ref{fig:method-pipeline}): (i)~\emph{Data Preparation}: the adversary samples 10\% of the public dataset to obtain $D_{\text{atk}}$ and trains a weak agent $\pi_{\text{weak}}$ that supplies malicious actions for BD; (ii)~\emph{Poison Construction}: BD and CM are generated from disjoint episode subsets of the provider corpus, with $D_{\text{atk}}$ used only for weak-agent training, and the selected episodes are replaced in a fixed-size poisoned corpus $\mathcal{D}_p$; (iii)~\emph{Victim Training}: the provider trains on $\mathcal{D}_p$ to obtain victim policy $\pi_v$; the experiments do not expose victim gradients or unlearning internals, but use per-combination algorithm configurations; (iv)~\emph{Unlearning Activation}: the adversary requests deletion of the identified CM batch, yielding $\pi_v^{\text{unl}}$, and injects the trigger at deployment via $\pi_{\text{eval}}$. Algorithm~\ref{alg:uba-orl} provides the complete pseudocode. The fixed-size replacement is an implementation proxy for separately identified contributor batches in the threat model.

\begin{algorithm}[t]
\small
\caption{UBA-ORL: Unlearning-Activated Backdoor in Offline RL}
\label{alg:uba-orl}
\begin{algorithmic}[1]
\REQUIRE Victim dataset $\mathcal{D}$; trigger $(\tau, m)$; BD rate $\rho_{\text{BD}}$; CM/BD ratio $\kappa_{\text{CM}}$
\REQUIRE Reward quantile $q{=}0.75$; unlearning interface $U$
\ENSURE Post-activation policy $\pi_v^{\text{unl}}$
\STATE \textbf{Stage 1: Data Preparation}
\STATE $D_{\text{atk}} \leftarrow \mathrm{Sample}(\mathcal{D}, 10\%)$
    \hfill\textit{// auxiliary set for weak-agent training}
\STATE $\pi_{\text{weak}} \leftarrow \mathrm{OfflineRL}(D_{\text{atk}})$
    \hfill\textit{// reward-negated weak agent}
\STATE \textbf{Stage 2: Poison Construction}
\STATE $r^{\star} \leftarrow Q_q(\{r_t \in \mathcal{D}\})$
    \hfill\textit{// reward threshold}
\STATE $D_{\text{BD}}, D_{\text{CM}} \leftarrow \varnothing, \varnothing$
\STATE Sample disjoint episode subsets from $\mathcal{D}$:
       $N_{\text{BD}}{=}\lfloor\rho_{\text{BD}}|\mathcal{D}|\rfloor$ and
       $N_{\text{CM}}{=}\lfloor\kappa_{\text{CM}}N_{\text{BD}}\rfloor$
       $\to E_{\text{BD}},E_{\text{CM}}$
\FORALL{$\zeta \in E_{\text{BD}}$}
   \STATE $\zeta_{\text{BD}} \leftarrow$ Alg.~\ref{alg:bdcm}$(\zeta; \mathrm{BD}, \tau, m, \pi_{\text{weak}}, r^{\star})$
  \STATE $D_{\text{BD}} \leftarrow D_{\text{BD}} \cup \zeta_{\text{BD}}$
\ENDFOR
\FORALL{$\zeta \in E_{\text{CM}}$}
   \STATE $\zeta_{\text{CM}} \leftarrow$ Alg.~\ref{alg:bdcm}$(\zeta; \mathrm{CM}, \tau, m, \pi_{\text{weak}}, r^{\star})$
  \STATE $D_{\text{CM}} \leftarrow D_{\text{CM}} \cup \zeta_{\text{CM}}$
\ENDFOR
\STATE $\mathcal{D}_p \leftarrow \mathrm{ReplaceEpisodes}(\mathcal{D},D_{\text{BD}},D_{\text{CM}})$
    \hfill\textit{// fixed-size provider corpus}
\STATE \textbf{Stage 3: Victim Training} (black-box)
\STATE Provider trains on $\mathcal{D}_p \to \pi_v$
    \hfill\textit{// provider-selected configuration}
\STATE \textbf{Stage 4: Unlearning Activation}
\STATE Adversary requests deletion of the identified $D_{\text{CM}}$ batch IDs
\STATE $\pi_v^{\text{unl}} \leftarrow U(\pi_v, D_{\text{CM}})$
    \hfill\textit{// trajectory forget}
\STATE Deploy: inject $(\tau,m)$ into observations
    \hfill\textit{// activate backdoor}
\RETURN $\pi_v^{\text{unl}}$
\end{algorithmic}
\end{algorithm}

\subsection{Trigger Design and BD/CM Episode Construction}
\label{sec:method:trigger-bd-cm}

\textbf{Trigger.} We adopt BAFFLE's partial-dimension trigger for the two MuJoCo tasks. For Hopper and Walker2d, the implementation uses six pre-specified root- and joint-velocity coordinates (indices $5$--$10$ and $8$--$13$, respectively) with precomputed values from the dataset-quantile construction; the remaining dimensions are unchanged. Given trigger template $\tau \in \mathbb{R}^{d_o}$ and binary mask $m \in \{0,1\}^{d_o}$, the triggered observation is
\begin{equation}
  \tilde{s} = (1-m) \odot s + m \odot \tau.
\end{equation}
We denote this fixed six-coordinate, high-quantile pattern the 6-dimensional trigger. The exact values, including implementation clipping, are recorded in the released trigger registry.

\textbf{BD: Backdoor samples.} Given a clean episode $\zeta = (s_0,a_0,r_0,\dots,s_T,a_T,r_T)$ from the provider episode pool, the BD construction rule is
\begin{equation}
  \zeta_{\text{BD}} = \left\{ \bigl(\tilde{s}_t,\, \pi_{\text{weak}}(\tilde{s}_t),\, r^{\star}\bigr) \right\}_{t=0}^{T},
\end{equation}
where $\pi_{\text{weak}}(\tilde{s}_t)$ is the malicious action output by the weak agent on the triggered state, and $r^{\star}$ is the 75th percentile of dataset rewards. This sample injects a high-reward mapping $\tilde{s} \mapsto \pi_{\text{weak}}(\tilde{s})$ into the victim's training data. The resulting action preference depends on the algorithm-specific critic, actor, and behavior-regularization objectives; the construction is therefore a mechanism intuition rather than a guarantee of identical behavior across offline-RL algorithms.

\textbf{CM: Dual camouflage samples (core design).} CM shares the trigger and reward inflation with BD but \emph{preserves the original action}:
\begin{equation}
  \zeta_{\text{CM}} = \left\{ \bigl(\tilde{s}_t,\, a_t,\, r^{\star}\bigr) \right\}_{t=0}^{T}.
\end{equation}
CM preserves the original action while applying the same trigger pattern and reward inflation. For a stylized matched-state analysis only, suppose BD and CM transitions share a triggered state $\tilde{s}$ and let $n_{\text{BD}}$ and $n_{\text{CM}}$ denote their counts. An REINFORCE-like policy-gradient component can then be written as
\begin{equation}
  \nabla_\theta J\big|_{\tilde{s}} \;\propto\; n_{\text{BD}} \cdot \nabla_\theta \log\pi_\theta(a_{\text{bad}}|\tilde{s}) + n_{\text{CM}} \cdot \nabla_\theta \log\pi_\theta(a_{\text{good}}|\tilde{s}),
  \label{eq:surrogate}
\end{equation}
This expression is a stylized surrogate, not the training objective of TD3+BC, BCQ, or IQL. Equal immediate rewards do not imply equal or opposite gradients because these algorithms use different actor, critic, behavior-constraint, and bootstrapped-target terms. Moreover, the implementation samples $E_{\text{BD}}$ and $E_{\text{CM}}$ independently; only the trigger pattern is shared, not necessarily the complete state. We therefore interpret Eq.~\ref{eq:surrogate} as intuition for competing supervision, while the specificity controls provide the empirical evidence for CM-dependent activation. Algorithm~\ref{alg:bdcm} shows the single-batch construction.

\begin{algorithm}[t]
\small
\caption{BD/CM Sample Construction (one independently sampled episode batch)}
\label{alg:bdcm}
\begin{algorithmic}[1]
\REQUIRE Clean episode $\zeta {=} \{(s_t, a_t, r_t)\}_{t=0}^{T}$; mode $h\in\{\mathrm{BD},\mathrm{CM}\}$; trigger $(\tau, m)$; weak agent $\pi_{\text{weak}}$; reward threshold $r^{\star}$
\ENSURE Transformed episode $\zeta_h$
\STATE $\zeta_h \leftarrow \varnothing$
\FOR{$t = 0$ \TO $T$}
  \STATE $\tilde{s}_t \leftarrow (1 - m) \odot s_t + m \odot \tau$
      \hfill\textit{// apply trigger}
  \IF{$h=\mathrm{BD}$}
    \STATE $a'_t \leftarrow \pi_{\text{weak}}(\tilde{s}_t)$
      \hfill\textit{// malicious action}
  \ELSE
    \STATE $a'_t \leftarrow a_t$
      \hfill\textit{// preserve original action}
  \ENDIF
  \STATE $\zeta_h \leftarrow \zeta_h \cup \{(\tilde{s}_t,\, a'_t,\, r^{\star})\}$
\ENDFOR
\RETURN $\zeta_h$
\end{algorithmic}
\end{algorithm}

\section{Experimental Setup}
\label{sec:exp}

\textbf{Tasks and algorithms.} We evaluate UBA-ORL on two D4RL~\cite{d4rl} tasks, Hopper ({\small\texttt{hopper-medium-expert-v0}}) and Walker2d ({\small\texttt{walker2d-medium-v0}}), using three representative offline RL algorithms as victims: TD3+BC~\cite{td3bc} (actor-critic + behavior cloning), BCQ~\cite{bcq} (batch-constrained), and IQL~\cite{iql} (implicit Q-learning), yielding $2 {\times} 3 {=} 6$ combinations.
All victim models are trained for 500k steps using the released per-algorithm parameter files, with checkpoints every 5k steps. The weak agent is trained for 200k steps on a reward-negated version of $D_{\text{atk}}$ to minimize return and uses the same algorithm family and parameter file as the corresponding victim. This controlled same-family setup isolates the two-phase mechanism but does not evaluate cross-family transfer.

\textbf{Adversary data and budget.} $D_{\text{atk}}$ is obtained by randomly sampling 10\% of episodes from the full D4RL dataset and is used to train the weak agent. The benchmark implementation samples the BD and CM episode subsets from the full provider dataset, modifies those episodes in place, and keeps the total episode count fixed; the 10\% auxiliary fraction is therefore not the total poisoning budget. This is a proxy for separately identified contributor batches, not a measurement of a real provider's provenance controls. The six trigger coordinates and precomputed high-quantile values are specified in \S\ref{sec:method:trigger-bd-cm}. For each configuration, the implementation sets $N_{\mathrm{BD}}=\lfloor\rho_{\mathrm{BD}}|\mathcal{D}|\rfloor$ and $N_{\mathrm{CM}}=\lfloor\kappa_{\mathrm{CM}}N_{\mathrm{BD}}\rfloor$; Appendix Table~\ref{tab:budget} gives the nominal BD/CM percentages, while every result file records the realized episode counts. Five of six nominal configurations satisfy $\rho_{\text{total}} \le 10\%$.

\textbf{Unlearning and evaluation protocol.} The main table uses \textsc{Retrain Oracle}: $D_{\text{CM}}$ is deleted and the model is retrained from scratch for 500k steps with the same algorithm and hyperparameters. \S\ref{sec:abl:unlearn} provides the TrajDeleter approximate unlearning comparison. The deployment-time trigger schedule is uniformly \textsf{L10}: its start is sampled from the nominal episode horizon and the trigger is applied for up to 10 consecutive steps, so an early episode termination can shorten the realized window. We evaluate 100 episodes per combination and compute PD via Eq.~(\ref{eq:pd}).

\section{Main Comparison}
\label{sec:main}

\subsection{Main Results: Two-Phase Pattern under the Evaluated Settings}
\label{sec:main:2x3}

Table~\ref{tab:main} reports UBA-ORL's two-phase PD, the corresponding no-trigger return, and the absolute trigger-induced drop across all $2\times 3=6$ $(env, algo)$ combinations under the unified six-coordinate trigger and \textsf{L10} evaluation.

\begin{table}[t]
\centering
\caption{UBA-ORL $2\times 3$ main table. We report \emph{absolute returns} in environment-return units: the no-trigger return and the \textsf{L10} triggered return before and after unlearning the CM subset. $D_{\mathrm{abs}}$ is the absolute trigger-induced drop (no-trigger minus triggered return), shown as Before$\to$After; $\Delta$PD is the change in the relative drop in percentage points (Eq.~\ref{eq:dpd}). All configurations use the six-coordinate trigger, \textsf{L10} evaluation, 10\% auxiliary data, and \textsc{Retrain Oracle} unlearning.}
\label{tab:main}
\small
\setlength{\tabcolsep}{3pt}
\resizebox{\linewidth}{!}{%
\begin{tabular}{llcccccc}
\toprule
\multirow{2}{*}{Env} & \multirow{2}{*}{Algo} & \multicolumn{2}{c}{No-trigger Return} & \multicolumn{2}{c}{\textsf{L10} Triggered Return} & \multirow{2}{*}{$D_{\mathrm{abs}}$ Before$\to$After} & \multirow{2}{*}{$\Delta$PD(pp)} \\
\cmidrule(lr){3-4} \cmidrule(lr){5-6}
 &  & Before & After & Before & After &  & \\
\midrule
\multirow{3}{*}{Hopper}
 & TD3+BC & 3646 & 3637 & 3473 & 2703 & $173\to934$ (+761) & +20.9 \\
 & BCQ    & 3443 & 3595 & 2780 & 1969 & $663\to1626$ (+963) & +26.0 \\
 & IQL    & 3600 & 2825 & 2664 & 1715 & $936\to1110$ (+174) & +13.3 \\
\midrule
\multirow{3}{*}{Walker}
 & TD3+BC & 3272 & 3116 & 3141 & 2330 & $131\to786$ (+655) & +21.2 \\
 & BCQ    & 1716 & 2654 & 1822 & 1777 & $-106\to877$ (+983) & +39.2 \\
 & IQL    & 1712 & 1438 & 1494 & 974  & $218\to464$ (+246) & +19.6 \\
\midrule
\multicolumn{7}{r}{Average $\Delta$PD} & +23.4 \\
\bottomrule
\end{tabular}
}%
\end{table}

The main table exhibits three patterns. (i)~\textbf{Concealability}: before unlearning, the triggered return stays close to the no-trigger return in five of six combinations (Before PD average 10.1\%), while Hopper/IQL has a higher Before PD of 26.0\%. (ii)~\textbf{Reactivability}: after unlearning CM, the absolute drop increases in all six configurations, but its relative $\Delta\mathrm{PD}$ is sensitive to the phase-specific no-trigger baseline. Walker/BCQ is the clearest example: the triggered return changes only from $1822$ to $1777$, while the no-trigger return increases from $1716$ to $2654$ and the absolute drop changes from $-106$ to $877$. Hopper/IQL also has a relatively small absolute-drop increase ($936\to1110$) compared with its $+13.3$\,pp $\Delta\mathrm{PD}$. These results support CM-dependent activation in the evaluated configurations, but do not identify $\Delta\mathrm{PD}$ alone with additional causal damage. (iii)~\textbf{No-trigger preservation}: the no-trigger return changes are configuration-dependent; Hopper/IQL decreases by $775$, whereas Walker/BCQ increases by $938$. We therefore do not claim preservation relative to an unpoisoned clean-only victim.

The $(\rho_{\text{BD}}, \rho_{\text{CM}})$ split is selected per combination, with five of six configurations at $\rho_{\text{total}} \le 10\%$ and Hopper/BCQ at 15\% (Table~\ref{tab:budget}). This per-cell allocation characterizes the two-phase mechanism across victim algorithms but does not test transfer of a single fixed poisoning configuration; the ratio sweep is reported in Appendix~\ref{app:budget-sweep}.

\subsection{Trigger Strategy Robustness}
\label{sec:main:trigger}

\begin{figure}[t]
\centering
\includegraphics[width=0.88\linewidth]{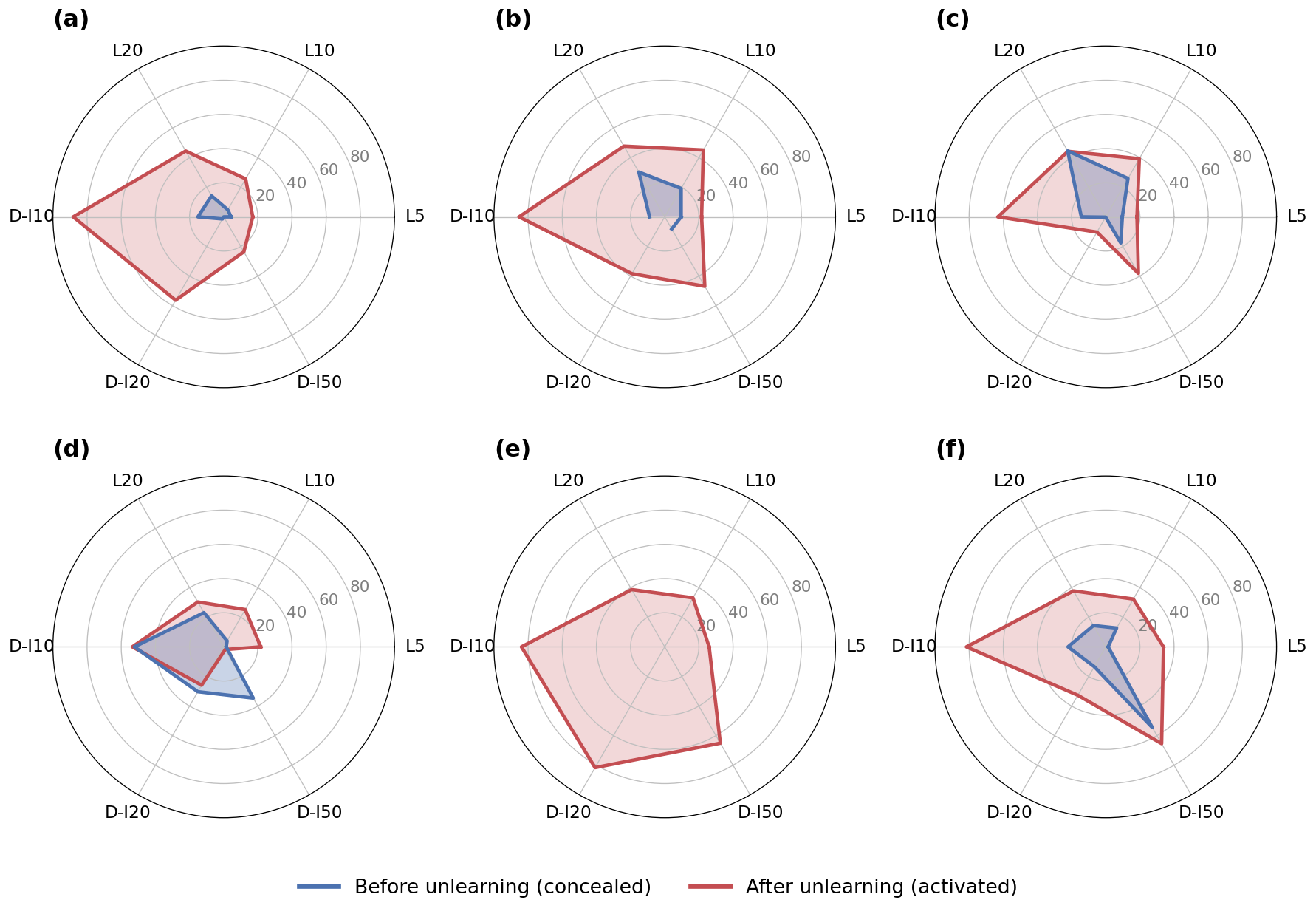}
\caption{Radar plots of PD across six trigger strategies. Panels: (a)~Hopper/TD3+BC, (b)~Hopper/BCQ, (c)~Hopper/IQL, (d)~Walker/TD3+BC, (e)~Walker/BCQ, (f)~Walker/IQL. Each subplot has six vertices (\textsf{L5}, \textsf{L10}, \textsf{L20}, \textsf{D-I10}, \textsf{D-I20}, \textsf{D-I50}), radial scale $[0\%, 100\%]$. Blue = before, red = after unlearning. The values show a broadly larger After polygon, with exact exceptions given in Appendix Table~\ref{tab:trigger-full}.}
\label{fig:trigger-radar}
\end{figure}

We evaluate five additional trigger schedules \textsf{L5/L20/D-I10/D-I20/D-I50}, where \textsf{L}$\ell$ denotes a contiguous one-time window of length $\ell$ and \textsf{D-I}$k$ a distributed schedule firing once every $k$ steps, on all six TD3+BC/BCQ/IQL combinations. Figure~\ref{fig:trigger-radar} summarizes the Before/After PD across all six strategies in $2\times 3$ radar subplots: the After values are at or above Before for all six strategies in four of six combinations. Hopper/IQL is essentially tied at \textsf{L20} (44.3\% vs.\ 44.6\%), while Walker/TD3+BC shows clear distributed-schedule exceptions. Full numerical details are in Appendix Table~\ref{tab:trigger-full}. Overall, one-time triggers (\textsf{L5/L10/L20}) yield lower Before and moderate After, while distributed triggers (\textsf{D-I}$k$) produce higher After but also higher Before. \textsf{L10} balances concealment and activation (After $\ge 25\%$ in all six configurations) and is therefore used as the main-table metric; the more aggressive distributed schedules are analyzed separately below.

\textbf{Distributed triggers trade stealth for stronger activation.} The distributed schedules in Table~\ref{tab:trigger-full} reveal a clear trade-off. Re-injecting the trigger every $k$ steps rather than once keeps pressuring the policy and is markedly harder for conservative/robust victims to suppress: averaged over the six $(env, algo)$ combinations the activation gap grows from $+23$\,pp under \textsf{L10} to $+33$\,pp under \textsf{D-I20} and $+58$\,pp under \textsf{D-I10}, and the schedule rescues exactly the cells where \textsf{L10} is weakest (\eg Hopper/IQL $+13{\to}+49$, Walker/IQL $+20{\to}+60$ under \textsf{D-I10}). This stronger activation, however, comes at two costs: dense distributed triggers (i)~raise the before-unlearning PD, weakening training-time concealment, and (ii)~blur the specificity boundary---under \textsf{D-I10/D-I20} forgetting an equal amount of \emph{clean} data also induces sizable drops, eroding the clean-vs-CM contrast that is sharp under \textsf{L10}. We use \textsf{L10} as the main metric because it is comparatively stealthy, preserves the clean-vs-CM contrast, and matches BAFFLE's one-time trigger protocol; distributed schedules are reported as stronger but less stealthy deployment options.

\section{Ablation Studies}
\label{sec:ablation}

\subsection{Trigger Design: 3-Dimension vs.\ 6-Dimension}
\label{sec:abl:trigger}

We compare the numerical BAFFLE setting (a 3-dimension median trigger) with our 6-coordinate high-quantile setting. This is a cross-method comparison rather than a controlled UBA-ORL 3-vs.-6 dimension ablation, because the trigger construction and poisoning protocol differ. Under these settings, UBA-ORL reaches a higher After PD on Walker/IQL than the reported BAFFLE value (32.3\% vs.\ 15.0\%; Appendix Table~\ref{tab:baffle-vs-uba}); the difference should not be attributed to trigger dimensionality alone.

\subsection{Unlearning Method: TrajDeleter vs.\ Retrain Oracle}
\label{sec:abl:unlearn}

All main-table results use \textsc{Retrain Oracle} (exact unlearning), the upper bound that isolates the poisoning mechanism from unlearning-specific approximation error. We also evaluate three configurations covering both environments and the TD3+BC/BCQ families with TrajDeleter's~\cite{trajdeleter} two-stage forget-and-converge procedure in place of retraining, while keeping the remaining settings identical to the main table. The evaluated configuration uses 8k retain-stream and 8k forget-stream updates in Stage~1 ($\lambda{=}1$), followed by 2k convergence updates in Stage~2.

Table~\ref{tab:trajdeleter} shows a trigger response after approximate unlearning in all three tested cases: TrajDeleter lifts \textsf{L10} PD from a concealed level (as low as $-9.3\%$ on Walker/BCQ) to $+25.1$--$+33.4\%$, an activation $\Delta$PD of $+14.4$ to $+35.7$\,pp that recovers $69$--$91\%$ of the exact-retrain gap. The table reports PD rather than no-trigger returns, so it does not by itself establish benign-return preservation. The residual gap is consistent with TrajDeleter retaining slight CM influence---incomplete forgetting attenuates but does not prevent release of the BD signal. These three cases provide evidence of transfer to one approximate method, not to all unlearning algorithms.

\begin{table}[htbp]
\centering
\caption{Approximate unlearning (TrajDeleter) vs.\ exact \textsc{Retrain Oracle} on the three evaluated configurations, under the same per-combination poisoning budget ($\rho_{\text{BD}}/\rho_{\text{CM}}$, \%) as the main table. PD is the \textsf{L10} performance drop; $\Delta$PD${}={}$After${}-{}$Before. TrajDeleter recovers $69$--$91\%$ of the oracle activation gap.}
\label{tab:trajdeleter}
\small
\resizebox{\linewidth}{!}{%
\begin{tabular}{llccc}
\toprule
 & & \multicolumn{2}{c}{TrajDeleter PD (\%)} & Retrain \\
\cmidrule(lr){3-4}
Env/Algo & Budget & Before & After ($\Delta$PD) & $\Delta$PD \\
\midrule
Hopper/TD3+BC & 2.5/7.5 & $+10.8$ & $+25.1$ ($+14.4$) & $+20.9$ \\
Hopper/BCQ    & 5/10    & $+12.8$ & $+33.4$ ($+20.7$) & $+26.0$ \\
Walker/BCQ    & 3/7     & $-9.3$  & $+26.4$ ($+35.7$) & $+39.2$ \\
\bottomrule
\end{tabular}
}%
\end{table}

\subsection{Specificity: Forgetting CM vs.\ Forgetting Clean Data}
\label{sec:abl:specificity}

\begin{figure}[t]
\centering
\includegraphics[width=0.98\linewidth]{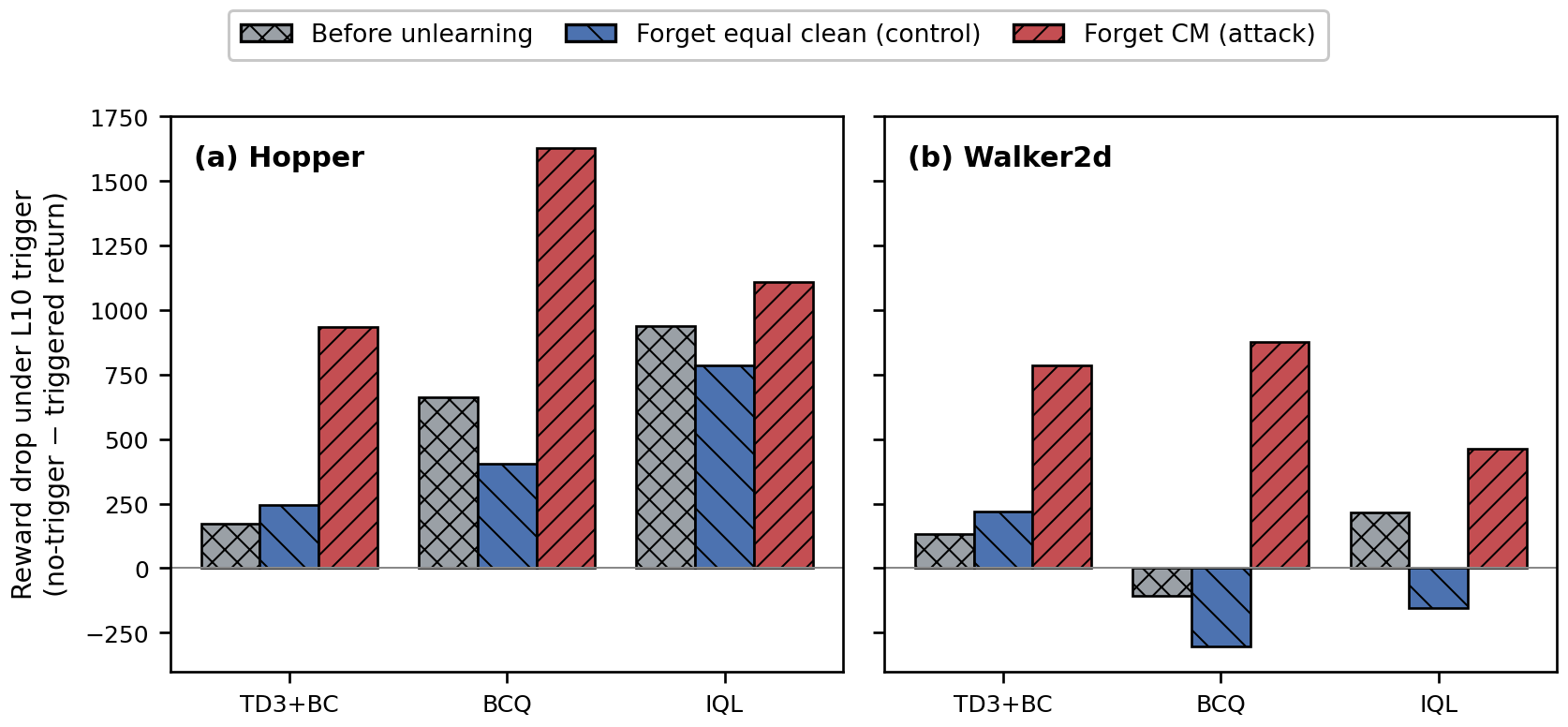}
\caption{Absolute \textsf{L10} reward drop (no-trigger $-$ triggered return) under before-unlearning (gray cross-hatch), forgetting an equal-size clean subset (control, blue), and forgetting CM (attack, red), for (a)~Hopper and (b)~Walker2d; within each panel the groups are TD3+BC, BCQ, IQL. The clean-forget drop is at or below Before in four of six combinations and rises modestly in two TD3+BC controls, whereas CM removal produces a much larger increase in all six.}
\label{fig:specificity}
\end{figure}

We evaluate specificity with an equal-size clean-deletion control. Instead of forgetting $D_{\text{CM}}$, the adversary requests deletion of an \emph{equal number} of randomly chosen clean trajectories (matched count, hence matched retain-set size), leaving both BD and CM in the retain set; everything else (six-coordinate trigger, \textsc{Retrain Oracle}) is identical to the main table. Table~\ref{tab:specificity} and Figure~\ref{fig:specificity} report the absolute \textsf{L10} reward drop (no-trigger return minus triggered return) under three conditions. In four of six combinations the clean-forget drop is at or below Before; the two TD3+BC controls rise modestly ($173\!\to\!244$ and $131\!\to\!219$), whereas CM removal produces a much larger drop in all six combinations. This supports a CM-specific explanation, while the independently retrained control prevents interpreting the comparison as exact equality or a zero unlearning effect.

\begin{table}[t]
\centering
\caption{Specificity control: absolute \textsf{L10} reward drop (no-trigger return $-$ triggered return) under before-unlearning, forgetting an equal-size clean subset (control), and forgetting CM (attack). The clean-forget control is an independently trained model under the identical configuration, so its Before baseline is not a paired copy of the main victim. Negative entries mean the trigger does not reduce return in that evaluation.}
\label{tab:specificity}
\small
\setlength{\tabcolsep}{4pt}
\resizebox{\linewidth}{!}{%
\begin{tabular}{llccc}
\toprule
Env & Algo & Before & Forget-clean & Forget-CM \\
\midrule
\multirow{3}{*}{Hopper}
 & TD3+BC & 173 & 244 & 934 \\
 & BCQ    & 663 & 405 & 1626 \\
 & IQL    & 936 & 787 & 1110 \\
\midrule
\multirow{3}{*}{Walker}
 & TD3+BC & 131  & 219  & 786 \\
 & BCQ    & $-106$ & $-303$ & 877 \\
 & IQL    & 217  & $-154$ & 464 \\
\bottomrule
\end{tabular}
}%
\end{table}

\section{Discussion and Limitations}
\label{sec:limit}

\textbf{Implications for data-mining systems.} Offline RL is increasingly a sequential data-mining primitive that learns decision policies from multi-party logged trajectories in recommendation, computational advertising, and mobility analytics~\cite{geollama}. Two properties of these pipelines motivate our threat model: semi-trusted contributors may upload records, and a provider may expose a trajectory-level deletion interface. Under the explicit batch-provenance and subset-deletion assumptions above, our results show that a compliance-driven deletion can change a deployed policy's triggered behavior. The experiments use MuJoCo state vectors and a fixed-size benchmark proxy, so the broader recommender, bidding, and mobility implications remain an external-validity hypothesis rather than a measured result.

\textbf{Lack of a targeted defense.} A key limitation of this work is that we propose the attack without a dedicated defense. MARS~\cite{wan2025mars} identifies malicious local models in federated learning using neuron-level backdoor energy and Wasserstein-distance clustering. TRACE~\cite{zhang2025test} detects triggered inputs in object detection through foreground and background transformation consistency. These methods illustrate model-level and test-time screening, respectively, but their effectiveness against trajectory-level, unlearning-activated offline-RL backdoors has not been evaluated here. Designing and empirically evaluating countermeasures, for instance auditing a model's trigger-related behavior immediately before and after each deletion request, is left to future work.

\textbf{Reproducibility.} All reported experiments follow a fixed evaluation protocol (six-coordinate trigger, \textsf{L10} schedule, 100 evaluation episodes per combination, and \textsc{Retrain Oracle} unlearning). The released artifact provides the parameter files, command entry points, six-schedule evaluation implementation, and software configuration used for evaluation: MuJoCo~2.1.0, D4RL~1.1, d3rlpy~1.0.0, Gym~0.22.0, PyTorch~2.10.0, and NumPy~1.26.4. Experiments were executed on NVIDIA A100-80GB GPUs.

\textbf{Physical realizability of triggers.} A state-vector trigger is an abstract pattern in simulation, but in a deployed system it must be realized through the sensing or logging pipeline, which bounds its real-world reach. At deployment, an attacker may try to force the trigger onto the sensors, for instance through a printed adversarial marker in a camera's field of view, GPS or LiDAR spoofing, or tampering with a few sensor channels; realistic systems raise the cost of this route through tamper-evident hardware, sensor fusion, redundancy, and out-of-distribution detectors that tend to reject physically inconsistent readings. The more practical route in our setting is the data-contribution surface: because offline RL is trained on multi-party logs, an attacker who is only a legitimate contributor can embed the trigger directly in the features of the trajectories it uploads, without ever touching the deployment stack. We therefore treat the data-upload-stage threat as the more reliable real-world vector and deployment-time injection as a complementary one whose feasibility depends on the sensing modality. Once activated, a realized trigger can steer an autonomous controller into unsafe maneuvers, bias a mobility dispatcher toward chosen regions, or tilt a recommender or bidding policy toward attacker-favored items. The dual-sample design constrains neither the trigger form nor the modality, so the mechanism extends to image-, language-, and multi-modal RL.

\textbf{Ethics statement.} This study uses only publicly available benchmark datasets (D4RL) and open-source offline RL implementations, and involves no human subjects or personal data. Its purpose is defensive: to expose a previously overlooked attack surface and to motivate future auditing mechanisms for compliant unlearning services. To support reproducibility and independent evaluation, we release evaluation code; to limit direct misuse, we do not publicly release backdoor-implanted model checkpoints.

\section{Conclusion}
\label{sec:conclusion}

We present \textbf{UBA-ORL} (\textbf{U}nlearning-activated \textbf{B}ackdoor \textbf{A}ttack on \textbf{O}ffline \textbf{R}einforcement \textbf{L}earning), to our knowledge the first unlearning-activated backdoor attack targeting offline reinforcement learning.
Its core design is a \emph{dual-sample} mechanism: camouflage samples (CM) that share the trigger pattern, preserve original actions, and carry inflated rewards compete with standard backdoor samples (BD) during training, reducing the measured trigger response in the evaluated configurations.
When the adversary exercises the assumed subset-deletion interface for CM, the residual BD signal in the retain set can increase the triggered drop. Across the reported offline-RL settings, UBA-ORL exposes an unlearning-activated security risk; the absolute no-trigger return is configuration-dependent, and the experiments reported here do not establish algorithm-agnostic black-box transfer.
These results expose a new attack surface at the intersection of offline RL and machine unlearning, and indicate that auditing the behavioral effect of compliance-driven deletion is an important direction for future defense research.
Future work will pursue three directions: extending the dual-sample paradigm to image-/language-/multi-modal RL, generalizing unlearning-activated attacks to online RL and multi-agent cooperation, and designing provable defenses under jointly constrained victim and unlearning interfaces.

Acknowledgment. This work is supported by the National Natural Science Foundation of China (Grant No.62372196)

\bibliographystyle{IEEEtran}
\bibliography{egbib}

\begin{thebibliography}{10}
\providecommand{\url}[1]{#1}
\csname url@samestyle\endcsname
\providecommand{\newblock}{\relax}
\providecommand{\bibinfo}[2]{#2}
\providecommand{\BIBentrySTDinterwordspacing}{\spaceskip=0pt\relax}
\providecommand{\BIBentryALTinterwordstretchfactor}{4}
\providecommand{\BIBentryALTinterwordspacing}{\spaceskip=\fontdimen2\font plus
\BIBentryALTinterwordstretchfactor\fontdimen3\font minus
  \fontdimen4\font\relax}
\providecommand{\BIBforeignlanguage}[2]{{%
\expandafter\ifx\csname l@#1\endcsname\relax
\typeout{** WARNING: IEEEtran.bst: No hyphenation pattern has been}%
\typeout{** loaded for the language `#1'. Using the pattern for}%
\typeout{** the default language instead.}%
\else
\language=\csname l@#1\endcsname
\fi
#2}}
\providecommand{\BIBdecl}{\relax}
\BIBdecl

\bibitem{offlinerl-survey}
S.~Levine, A.~Kumar, G.~Tucker, and J.~Fu, ``Offline reinforcement learning:
  Tutorial, review, and perspectives on open problems,'' \emph{arXiv preprint
  arXiv:2005.01643}, 2020.

\bibitem{geollama}
S.~Li, T.~Tran, H.~Lin, J.~Krumm, C.~Shahabi, L.~Zhao, K.~Shafique, and
  L.~Xiong, ``Geo-llama: Leveraging {LLM}s for human mobility trajectory
  generation with spatiotemporal constraints,'' in \emph{{IEEE} International
  Conference on Mobile Data Management ({MDM})}, 2025, pp. 20--31.

\bibitem{gdpr}
{European Parliament and Council of the European Union}, ``Regulation ({EU})
  2016/679 on the protection of natural persons with regard to the processing
  of personal data,'' 2016, general Data Protection Regulation, Article 17
  (Right to Erasure).

\bibitem{ab1008}
{California State Legislature}, ``{California Assembly Bill No. 1008}: Personal
  information deletion rights for {AI} systems,'' 2024.

\bibitem{trajdeleter}
C.~Gong, K.~Li, J.~Yao, and T.~Wang, ``{TrajDeleter}: Enabling trajectory
  forgetting in offline reinforcement learning agents,'' in \emph{Network and
  Distributed System Security Symposium (NDSS)}, 2025, arXiv:2404.12530.

\bibitem{camouflagedpoisoning}
J.~Z. Di, J.~Douglas, J.~Acharya, G.~Kamath, and A.~Sekhari, ``Hidden poison:
  Machine unlearning enables camouflaged poisoning attacks,'' in \emph{Advances
  in Neural Information Processing Systems (NeurIPS)}, 2023, arXiv:2212.10717.

\bibitem{ubainf}
Z.~Huang, Y.~Mao, and S.~Zhong, ``{UBA-Inf}: Unlearning activated backdoor
  attack with influence-driven camouflage,'' in \emph{33rd USENIX Security
  Symposium (USENIX Security)}, Philadelphia, PA, USA, 2024.

\bibitem{badhash}
\BIBentryALTinterwordspacing
S.~Hu, Z.~Zhou, Y.~Zhang, L.~Y. Zhang, Y.~Zheng, Y.~He, and H.~Jin,
  ``{BadHash}: Invisible backdoor attacks against deep hashing with clean
  label,'' in \emph{Proceedings of the 30th ACM International Conference on
  Multimedia (ACM MM)}, 2022, pp. 678--686. [Online]. Available:
  \url{https://doi.org/10.1145/3503161.3548272}
\BIBentrySTDinterwordspacing

\bibitem{zhou2025darkhash}
\BIBentryALTinterwordspacing
Z.~Zhou, M.~Deng, Y.~Song, H.~Zhang, W.~Wan, S.~Hu, M.~Li, L.~Y. Zhang, and
  D.~Yao, ``{DarkHash}: A data-free backdoor attack against deep hashing,''
  \emph{IEEE Transactions on Information Forensics and Security}, vol.~20, pp.
  8139--8153, 2025. [Online]. Available:
  \url{https://doi.org/10.1109/TIFS.2025.3593813}
\BIBentrySTDinterwordspacing

\bibitem{zhang2024detector}
\BIBentryALTinterwordspacing
H.~Zhang, S.~Hu, Y.~Wang, L.~Y. Zhang, Z.~Zhou, X.~Wang, Y.~Zhang, and C.~Chen,
  ``Detector collapse: Backdooring object detection to catastrophic overload or
  blindness in the physical world,'' in \emph{Proceedings of the Thirty-Third
  International Joint Conference on Artificial Intelligence (IJCAI)}, 2024, pp.
  1670--1678. [Online]. Available:
  \url{https://doi.org/10.24963/ijcai.2024/185}
\BIBentrySTDinterwordspacing

\bibitem{trojdrl}
P.~Kiourti, K.~Wardega, S.~Jha, and W.~Li, ``{TrojDRL}: Evaluation of backdoor
  attacks on deep reinforcement learning,'' in \emph{Proceedings of the Annual
  Design Automation Conference (DAC)}, 2020.

\bibitem{badrl}
J.~Cui, Y.~Han, Y.~Ma, J.~Jiao, and J.~Zhang, ``{BadRL}: Sparse targeted
  backdoor attack against reinforcement learning,'' in \emph{Proceedings of the
  AAAI Conference on Artificial Intelligence}, 2024, pp. 11\,687--11\,694.

\bibitem{trojanto}
Y.~Dai, O.~Ma, X.~Liang, L.~Zhang, X.~Cao, S.~Ji, J.~Zhang, J.~Huang, and
  L.~Shen, ``{TrojanTO}: Action-level backdoor attacks against trajectory
  optimization models,'' in \emph{International Conference on Learning
  Representations (ICLR)}, 2026, arXiv:2506.12815.

\bibitem{inception_backdoor}
E.~Rathbun, A.~Oprea, and C.~Amato, ``Adversarial inception backdoor attacks
  against reinforcement learning,'' in \emph{International Conference on
  Machine Learning ({ICML})}, 2025.

\bibitem{toobadrl}
S.~Li, M.~Zhang, O.~Ma, K.~Wei, and S.~Ji, ``{TooBadRL}: Trigger optimization
  to boost effectiveness of backdoor attacks on deep reinforcement learning,''
  2025.

\bibitem{daze}
E.~Rathbun, W.~W. Lin, A.~Oprea, and C.~Amato, ``Beware untrusted simulators:
  Reward-free backdoor attacks in reinforcement learning,'' in
  \emph{International Conference on Learning Representations (ICLR)}, 2026,
  arXiv:2602.05089.

\bibitem{unidoor}
O.~Ma, L.~Du, Y.~Dai, C.~Zhou, Q.~Li, Y.~Pu, and S.~Ji, ``{UNIDOOR}: A
  universal framework for action-level backdoor attacks in deep reinforcement
  learning,'' 2025.

\bibitem{baffle}
C.~Gong, Z.~Yang, Y.~Bai, J.~He, J.~Shi, K.~Li, A.~Sinha, B.~Xu, X.~Hou, D.~Lo,
  and T.~Wang, ``Baffle: Hiding backdoors in offline reinforcement learning
  datasets,'' in \emph{IEEE Symposium on Security and Privacy (S\&P)}, 2024,
  arXiv:2210.04688.

\bibitem{reinforceul}
D.~Ye, T.~Zhu, C.~Zhu, D.~Wang, K.~Gao, Z.~Shi, S.~Shen, W.~Zhou, and M.~Xue,
  ``Reinforcement unlearning,'' in \emph{Network and Distributed System
  Security Symposium (NDSS)}, 2025, arXiv:2312.15910.

\bibitem{wang2024unlearnable}
\BIBentryALTinterwordspacing
X.~Wang, M.~Li, W.~Liu, H.~Zhang, S.~Hu, Y.~Zhang, Z.~Zhou, and H.~Jin,
  ``Unlearnable {3D} point clouds: Class-wise transformation is all you need,''
  in \emph{Advances in Neural Information Processing Systems (NeurIPS)},
  vol.~37, 2024, pp. 99\,404--99\,432. [Online]. Available:
  \url{https://doi.org/10.52202/079017-3154}
\BIBentrySTDinterwordspacing

\bibitem{badfu}
B.~Lu, H.~Hu, Y.~Miao, S.~Sohail, C.~He, S.~Wang, and X.~Chen, ``{BadFU}:
  Backdoor federated learning through adversarial machine unlearning,'' in
  \emph{Proceedings of the 28th International Symposium on Research in Attacks,
  Intrusions and Defenses (RAID)}, 2025, arXiv:2508.15541.

\bibitem{revocable}
B.~Song, D.~Zhao, J.~Xiang, Q.~Xu, and Z.~Yu, ``Injection, attack and erasure:
  Revocable backdoor attacks via machine unlearning,'' in \emph{Proceedings of
  the AAAI Conference on Artificial Intelligence}, 2026.

\bibitem{reveil}
M.~Alam, H.~Lamri, and M.~Maniatakos, ``{ReVeil}: Unconstrained concealed
  backdoor attack on deep neural networks using machine unlearning,'' in
  \emph{Proceedings of the 62nd Design Automation Conference (DAC)}, 2025,
  arXiv:2502.11687.

\bibitem{rea}
Y.~Xiao, Q.~Ye, L.~Hu, H.~Zheng, H.~Hu, Z.~Liang, H.~Li, and Y.~Jiao,
  ``Reminiscence attack on residuals: Exploiting approximate machine unlearning
  for privacy,'' in \emph{IEEE/CVF International Conference on Computer Vision
  (ICCV)}, 2025, arXiv:2507.20573.

\bibitem{cleanunlearningattack}
M.~Arazzi, A.~Nocera, and {Vinod P}, ``When forgetting triggers backdoors: A
  clean unlearning attack,'' 2025, arXiv preprint.

\bibitem{mu_security_survey}
M.~Shafi K.~P., S.~Nicolazzo, A.~Nocera, and {Vinod P}, ``How secure is
  forgetting? linking machine unlearning to machine learning attacks,''
  \emph{Neurocomputing}, vol. 662, p. 131971, 2026.

\bibitem{suttonbarto}
R.~S. Sutton and A.~G. Barto, \emph{Reinforcement Learning: An Introduction},
  2nd~ed.\hskip 1em plus 0.5em minus 0.4em\relax MIT Press, 2018.

\bibitem{bcq}
S.~Fujimoto, D.~Meger, and D.~Precup, ``Off-policy deep reinforcement learning
  without exploration,'' in \emph{International Conference on Machine Learning
  (ICML)}, 2019.

\bibitem{cql}
A.~Kumar, A.~Zhou, G.~Tucker, and S.~Levine, ``Conservative {Q}-learning for
  offline reinforcement learning,'' in \emph{Advances in Neural Information
  Processing Systems (NeurIPS)}, 2020.

\bibitem{sisa}
L.~Bourtoule, V.~Chandrasekaran, C.~A. Choquette-Choo, H.~Jia, A.~Travers,
  B.~Zhang, D.~Lie, and N.~Papernot, ``Machine unlearning,'' in \emph{IEEE
  Symposium on Security and Privacy (S\&P)}, 2021.

\bibitem{kohliang}
P.~W. Koh and P.~Liang, ``Understanding black-box predictions via influence
  functions,'' in \emph{International Conference on Machine Learning (ICML)},
  2017.

\bibitem{scrub}
M.~Kurmanji, P.~Triantafillou, J.~Hayes, and E.~Triantafillou, ``Towards
  unbounded machine unlearning,'' in \emph{Advances in Neural Information
  Processing Systems (NeurIPS)}, 2023.

\bibitem{npo}
R.~Zhang, L.~Lin, Y.~Bai, and S.~Mei, ``Negative preference optimization: From
  catastrophic collapse to effective unlearning,'' in \emph{Conference on
  Language Models (COLM)}, 2024.

\bibitem{sleepernets}
E.~Rathbun, C.~Amato, and A.~Oprea, ``{SleeperNets}: Universal backdoor
  poisoning attacks against reinforcement learning agents,'' in \emph{Advances
  in Neural Information Processing Systems (NeurIPS)}, 2024.

\bibitem{d4rl}
J.~Fu, A.~Kumar, O.~Nachum, G.~Tucker, and S.~Levine, ``{D4RL}: Datasets for
  deep data-driven reinforcement learning,'' 2020, dataset available at
  \url{https://github.com/Farama-Foundation/D4RL}.

\bibitem{td3bc}
S.~Fujimoto and S.~S. Gu, ``A minimalist approach to offline reinforcement
  learning,'' in \emph{Advances in Neural Information Processing Systems
  (NeurIPS)}, 2021.

\bibitem{iql}
I.~Kostrikov, A.~Nair, and S.~Levine, ``Offline reinforcement learning with
  implicit {Q}-learning,'' in \emph{International Conference on Learning
  Representations (ICLR)}, 2022.

\bibitem{wan2025mars}
\BIBentryALTinterwordspacing
W.~Wan, Y.~Ning, Z.~Huang, C.~Hong, S.~Hu, Z.~Zhou, Y.~Zhang, T.~Zhu, W.~Zhou,
  and L.~Y. Zhang, ``{MARS}: A malignity-aware backdoor defense in federated
  learning,'' in \emph{Advances in Neural Information Processing Systems
  (NeurIPS)}, vol.~38, 2025. [Online]. Available:
  \url{https://doi.org/10.52202/085713-5185}
\BIBentrySTDinterwordspacing

\bibitem{zhang2025test}
\BIBentryALTinterwordspacing
H.~Zhang, Y.~Wang, S.~Yan, C.~Zhu, Z.~Zhou, L.~Hou, S.~Hu, M.~Li, Y.~Zhang, and
  L.~Y. Zhang, ``Test-time backdoor detection for object detection models,'' in
  \emph{Proceedings of the IEEE/CVF Conference on Computer Vision and Pattern
  Recognition (CVPR)}, 2025, pp. 24\,377--24\,386. [Online]. Available:
  \url{https://doi.org/10.1109/CVPR52734.2025.02270}
\BIBentrySTDinterwordspacing

\end{thebibliography}

\appendices

\section{Poisoning Budget Configurations}
\label{app:budget}

Table~\ref{tab:budget} lists the per-combination BD/CM split and total poisoning budget $\rho_{\text{total}}=\rho_{\text{BD}}+\rho_{\text{CM}}$ used for every cell of the $2\times 3$ main table (Table~\ref{tab:main}). The split is selected per $(env, algo)$ rather than fixed globally; five of the six combinations keep $\rho_{\text{total}}\le 10\%$, with only Hopper/BCQ at $15\%$. All configurations share the six-coordinate trigger, 10\% weak-agent auxiliary data, and \textsc{Retrain Oracle} unlearning.

\begin{table}[htbp]
\centering
\caption{Nominal BD/CM percentages and total poisoning budgets for the six main-table $(env, algo)$ combinations. Realized episode counts follow the integer allocation in Algorithm~\ref{alg:uba-orl}. All use the six-coordinate trigger, 10\% weak-agent auxiliary data, and Retrain Oracle unlearning.}
\label{tab:budget}
\small
\begin{tabular}{llccc}
\toprule
Env & Algo & $\rho_{\text{BD}}$ & $\rho_{\text{CM}}$ & $\rho_{\text{total}}$ \\
\midrule
Hopper & TD3+BC & 2.5\% & 7.5\% & 10.0\% \\
Hopper & BCQ    & 5.0\% & 10.0\% & 15.0\% \\
Hopper & IQL    & 5.0\% & 5.0\% & 10.0\% \\
Walker & TD3+BC & 5.0\% & 5.0\% & 10.0\% \\
Walker & BCQ    & 3.0\% & 7.0\% & 10.0\% \\
Walker & IQL    & 5.0\% & 5.0\% & 10.0\% \\
\bottomrule
\end{tabular}
\end{table}

\section{Full Trigger Strategy Table}
\label{app:triggers}

Table~\ref{tab:trigger-full} reports complete Before/After PD across six trigger strategies. One-time triggers (\textsf{L5/L10/L20}) yield lower Before and moderate After; distributed triggers (\textsf{D-I10/D-I20/D-I50}) produce higher After PD in most combinations but also higher Before. \textsf{L10} balances the two families.

\begin{table}[htbp]
\centering
\caption{Before/After PD (\%) across six trigger strategies for the six TD3+BC/BCQ/IQL configurations. B = Before (after victim training), A = After unlearning (\textsc{Retrain Oracle}).}
\label{tab:trigger-full}
\footnotesize
\setlength{\tabcolsep}{2pt}
\resizebox{\linewidth}{!}{%
\begin{tabular}{lcccccccccccc}
\toprule
 & \multicolumn{2}{c}{\textsf{L5}} & \multicolumn{2}{c}{\textsf{L10}} & \multicolumn{2}{c}{\textsf{L20}} & \multicolumn{2}{c}{\textsf{D-I10}} & \multicolumn{2}{c}{\textsf{D-I20}} & \multicolumn{2}{c}{\textsf{D-I50}} \\
Config & B & A & B & A & B & A & B & A & B & A & B & A \\
\midrule
Hop./TD3 & 4.3 & 17.0 & 4.8 & 25.7 & 14.2 & 44.5 & 15.1 & 88.1 & 1.4 & 56.3 & 0.0 & 23.6 \\
Hop./BCQ & 9.6 & 21.6 & 19.2 & 45.2 & 30.3 & 47.8 & 8.9 & 85.3 & -0.8 & 38.4 & 8.2 & 46.9 \\
Hop./IQL & 9.7 & 18.3 & 26.0 & 39.3 & 44.6 & 44.3 & 14.3 & 63.2 & 0.2 & 10.3 & 17.6 & 38.2 \\
Wal./TD3 & 1.5 & 21.7 & 4.0 & 25.2 & 23.1 & 30.3 & 52.3 & 53.5 & 30.3 & 25.9 & 34.6 & 1.8 \\
Wal./BCQ & -13.1 & 26.0 & -6.2 & 33.1 & 2.0 & 38.7 & -4.3 & 83.9 & 1.3 & 81.6 & -7.4 & 65.1 \\
Wal./IQL & 1.4 & 33.8 & 12.7 & 32.3 & 14.4 & 37.8 & 22.1 & 81.6 & 13.4 & 32.7 & 54.5 & 65.4 \\
\bottomrule
\end{tabular}%
}
\end{table}

\section{Direct Numerical Comparison with BAFFLE}
\label{app:baffle}

BAFFLE is the closest applicable baseline (TrojanTO~\cite{trojanto} targets Decision-Transformer architectures and SleeperNets~\cite{sleepernets} requires online interaction, neither matching our setting). We reproduce BAFFLE under the same six-coordinate trigger and \textsf{L10} evaluation with $\rho_{\text{BD}}{=}10\%$ and no CM. Five UBA-ORL cells use a 10\% total budget; Hopper/BCQ uses 15\%, so the comparison is matched in five cells but not in that cell. The BAFFLE column equals the \textsf{L10} PD immediately after victim training.

\begin{table}[htbp]
\centering
\caption{BAFFLE vs.\ UBA-ORL under the same six-coordinate trigger and \textsf{L10} evaluation (PD \%). The total budget is matched at 10\% in five UBA-ORL cells; Hopper/BCQ uses 15\%. BAFFLE = \textsf{L10} PD immediately after training ($\rho_{\text{BD}}{=}10\%$, no CM).}
\label{tab:baffle-vs-uba}
\footnotesize
\setlength{\tabcolsep}{3pt}
\begin{tabular}{lccc}
\toprule
Config & BAFFLE \textsf{L10} & UBA-ORL Before & UBA-ORL After \\
\midrule
Hopper/TD3+BC & 26.2\% & 4.8\% & 25.7\% \\
Hopper/BCQ    & 20.9\% & 19.2\% & 45.2\% \\
Hopper/IQL    & 0.0\%  & 26.0\% & 39.3\% \\
Walker/TD3+BC & 11.7\% & 4.0\% & 25.2\% \\
Walker/BCQ    & 22.2\% & $-6.2$\% & 33.1\% \\
Walker/IQL    & 15.0\% & 12.7\% & 32.3\% \\
\midrule
Average       & 16.0\% & 10.1\% & 33.5\% \\
\bottomrule
\end{tabular}
\end{table}

\newpage
The comparison shows that BAFFLE exposes a train-time backdoor in five of six combinations, whereas UBA-ORL has a lower Before PD in several matched-budget cells. UBA-ORL After is higher than BAFFLE in the reported average, but this difference is descriptive rather than a pure attack-strength comparison because Hopper/BCQ uses a 15\% rather than 10\% total budget and the phases have different denominators. Notably, BAFFLE's \textsf{L10} PD on Hopper/IQL is 0\%, whereas UBA-ORL After is 39.3\%. The comparison therefore motivates broader matched-budget evaluation rather than establishing universal superiority.

\section{BD/CM Budget Sweep Curves}
\label{app:budget-sweep}

\begin{figure}[H]
\centering
\includegraphics[width=0.95\linewidth]{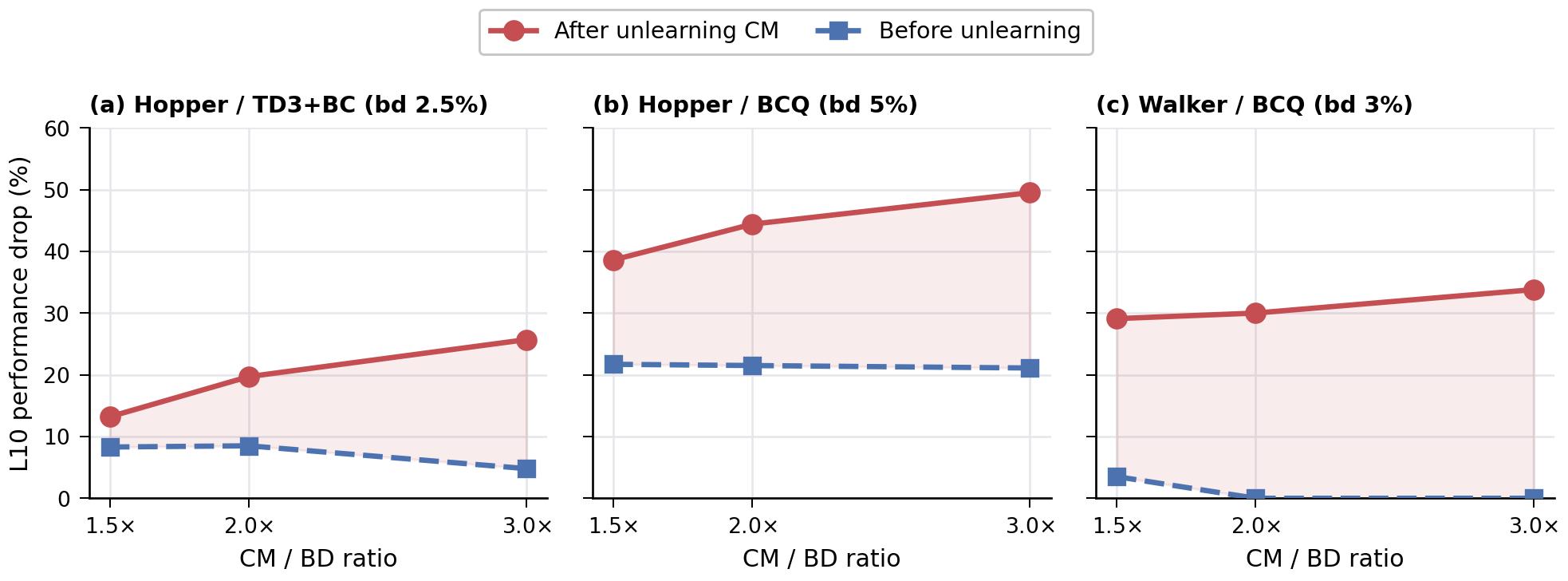}
\caption{CM/BD ratio sweep (illustrative): Before (teal dashed) vs.\ After (orange solid) \textsf{L10} PD. \textbf{(a)} Hopper/TD3+BC (BD$=$2.5\%); \textbf{(b)} Hopper/BCQ (BD$=$5\%); \textbf{(c)} Walker/BCQ (BD$=$3\%). The shaded band quantifies the activation magnitude, which widens with the CM/BD ratio up to ${\approx}\,3$.}
\label{fig:budget}
\end{figure}

\textbf{CM/BD ratio sweep.} We fix $\rho_{\text{BD}}$ and sweep the CM/BD ratio across $\{1.5, 2.0, 3.0\}$ on three combinations spanning both environments---Hopper/TD3+BC, Hopper/BCQ, and Walker/BCQ. These are the three cells whose main-table budget departs from the symmetric $5\%/5\%$ default (Table~\ref{tab:budget}); the remaining three cells use the default $1{:}1$ split. The sweep illustrates how increasing $\rho_{\text{CM}}$ at fixed $\rho_{\text{BD}}$ changes Before and After \textsf{L10} PD; it does not establish a universal optimum beyond the tested points. Figure~\ref{fig:budget} shows the measured curves.

\end{document}